\documentclass[superscriptaddress,aps,prl,reprint,twocolumn,amsmath,amssymb]{revtex4-2}
\usepackage{graphicx}
\usepackage{hyperref}
\usepackage{amssymb}
\usepackage{slashed}
\usepackage{dcolumn}
\usepackage{amsmath}
\usepackage{bm}% bold math
\usepackage{colordvi}
\usepackage{algorithm}
\usepackage{algpseudocode}
\usepackage{multirow}
\usepackage{titlesec}
\usepackage{newtxtext,newtxmath}
\usepackage{dsfont}
\usepackage{xcolor}
\usepackage{mathbbol}

\usepackage{hyperref}% add hypertext capabilities
\hypersetup{
    colorlinks=true,
    linkcolor=blue,
    citecolor=blue,     
    urlcolor=blue,
}% Added by user
\allowdisplaybreaks% Added by user

\usepackage{mathrsfs}
\makeatletter

\newcommand{\Rmnum}[1]{\expandafter\@slowromancap\romannumeral #1@}
\makeatother

\begin{document}

\preprint{APS/123-QED}

\title{Hidden Staggered Superconducting Order in V$_2$Se$_2$O-Family Altermagnets}% Force line breaks with \\

\author{Chun Yu Wan}
\affiliation{Department of Physics, The Hong Kong University of Science and Technology, Hong Kong, China}
\author{Xiangxi Zhang}
\affiliation{Department of Physics, The Hong Kong University of Science and Technology, Hong Kong, China}
\author{Fei Yang}
 \email{flyyangphys@ust.hk}
\affiliation{Department of Physics, The Hong Kong University of Science and Technology, Hong Kong, China}
\author{Junwei Liu}%
 \email{liuj@ust.hk}
\affiliation{Department of Physics, The Hong Kong University of Science and Technology, Hong Kong, China}
    \affiliation{IAS Center for Quantum Matter, Hong Kong University of Science and Technology, Clear Water Bay, Hong Kong SAR, China}

\date{\today}% It is always \today, today,

\begin{abstract}
$\eta$ pairing describes a staggered superconducting condensate whose order parameter changes sign between neighboring unit cells while maintaining a uniform magnitude. Nearly four decades after its theoretical proposal by C.~N.~Yang, a  material platform for stabilizing $\eta$ pairing has remained elusive. Here, we reveal that altermagnetism, despite being pair breaking for conventional BCS pairing,  naturally favors $\eta$-like pairing in V$_2$Se$_2$O-family altermagnets. Their valley-contrasting spin structure gives rise to oppositely spin-polarized, symmetry-related valleys centered at $X=(\pi,0)$ and $Y=(0,\pi)$, which form spin-singlet pairing partners separated by the commensurate momentum $\mathbf Q=(\pi,\pi)$, thereby favoring an $\eta$-like condensate. Self-consistent calculations demonstrate that the resulting $\eta$-like state remains robust over broad variations of chemical potential, hopping parameters, and Fermi-surface topology. To detect this hidden staggered order, we propose a phase-sensitive Josephson measurement with a uniform superconductor, where destructive interference suppresses single-Cooper-pair tunneling while preserving two-Cooper-pair processes. This constitutes a charge-$4e$-like Josephson response despite a bulk condensate composed of charge-$2e$ Cooper pairs, providing a direct phase-sensitive signature of the otherwise elusive $\eta$-like condensate in V$_2$Se$_2$O-family altermagnets.

\end{abstract}

\maketitle

%\tableofcontents

{\sl Introduction.---}Altermagnetism has recently expanded the landscape of collinear magnetism beyond the conventional dichotomy between ferromagnets and antiferromagnets~\cite{PhysRevX.12.031042,PhysRevX.12.040501,bai2024altermagnetism,song2025altermagnets}.
Altermagnets possess fully compensated magnetic order, while their opposite-spin sublattices are connected by crystal rotations or mirror operations, giving rise to crystal-symmetry-related, nonrelativistic spin-momentum locking~\cite{PhysRevX.12.031042,PhysRevX.12.040501,naka2019spin,smejkalCrystalTimereversalSymmetry2020,ahn2019antiferromagnetism, yuanGiantMomentumdependentSpin2020, hayamiMomentumDependentSpinSplitting2019, ma2021multifunctional, mazinPredictionUnconventionalMagnetism2021, liuSpinGroupSymmetryMagnetic2022}. Consequently, their electronic bands exhibit alternating spin polarization in momentum space despite a vanishing macroscopic magnetization~\cite{hu2025catalog}. This distinctive electronic structure enables phenomena inaccessible to either ferromagnets or conventional antiferromagnets~\cite{PhysRevB.75.115103}, making altermagnets promising platforms for studying novel spintronics~\cite{PhysRevLett.126.127701,PhysRevLett.130.216701,han2024electrical,sun2025spin,zhou2025manipulation,jungwirth2026altermagnetic} and quantum transport~\cite{ma2021multifunctional,feng2022anomalous,liao2024separation,sato2024altermagnetic,fu2025all}, as well as a fertile ground for unconventional electronic instabilities~\cite{fukaya2025superconducting,fukaya2025josephson,PhysRevB.111.184515,PhysRevLett.131.076003,PhysRevB.108.054511,dlpb-gfct}.

Among various electronic instabilities, superconductivity is particularly sensitive to the momentum-dependent spin structure of systems~\cite{Yang2017}.
A conventional spin-singlet condensate pairs opposite-spin states at opposite momenta, whereas the altermagnetism introduces an energetic mismatch between these pairing partners and therefore acts as a pair-breaking field for conventional BCS superconductivity~\cite{mazin2025notes,chakraborty2025constraints}.
This competition has motivated proposals for unconventional triplet $p$-wave pairing~\cite{PhysRevB.111.054501}, FFLO superconducting (SC) state~\cite{Zhang2024,whm1-76jc,6wxh-p4mc,PhysRevResearch.5.043171,rqp1-jtcb,mukasa2025finite, 3k12-2467, cvq6-b1nf}, spin-polarized supercurrents~\cite{52wh-1z5y}, SC diode effects~\cite{rqp1-jtcb}, and topological superconductivity~\cite{PhysRevB.108.184505,7h2t-wmyr} in altermagnets, substantially broadening the landscape of SC phases, transport phenomena, and device functionalities. 

Most existing theories of superconductivity in altermagnets 
have focused on systems with $\Gamma$-centered Fermi surfaces, whose
spin splitting changes sign across symmetry-enforced nodal
directions~\cite{Zhang2024,whm1-76jc,6wxh-p4mc,PhysRevResearch.5.043171,rqp1-jtcb,mukasa2025finite, 3k12-2467, cvq6-b1nf}.
A qualitatively different electronic structure emerges in the
V$_2$Se$_2$O family, a class of layered tetragonal compounds built from
V$_2$O square-lattice planes separated by chalcogen and spacer
layers~\cite{lin2018structure,liu2025physical,jiang2025metallic,zhang2025crystal,hu2026observation,sun2026emergent,li2026pressure}.
Within each V$_2$O plane, the V moments form a checkerboard
antiferromagnetic texture~\cite{ma2021multifunctional,xu2025chemical,cheng2026realistic}.  This in-plane magnetic structure is characterized by the spin-group symmetries
$[C_2\Vert C_{4z}]$ and $[C_2\Vert M_{xy}]$, where $C_2$ denotes a
$\pi$ spin rotation that reverses the spin polarization, while
$C_{4z}$ and $M_{xy}$ exchange the two magnetic sublattices and
interchange the zone-boundary points $X=(\pi,0)$ and $Y=(0,\pi)$.
Consequently, in the V$_2$Se$_2$O family, the $X$ and $Y$ valleys are symmetry-related yet oppositely
spin polarized, yielding strongly spin-polarized, nearly nodeless low-energy
states that are qualitatively distinct from the nodal spin splitting
characteristic of the $\Gamma$-centered Fermi surfaces considered previously.

Here we reveal that the spin-valley structure of V$_2$Se$_2$O-family altermagnets, while pair breaking for conventional BCS pairing, naturally favors staggered superconductivity, or $\eta$-like pairing. 
Originally introduced by C. N. Yang nearly four decades ago as exact many-body eigenstates of the Hubbard model~\cite{PhysRevLett.63.2144},  $\eta$-paired states carry the commensurate momentum $\mathbf Q=(\pi,\pi)$ and exhibit off-diagonal long-range order. Nevertheless, their realization in a material platform has remained elusive. The oppositely polarized $X$ and $Y$ valleys in V$_2$Se$_2$O-family altermagnets~\cite{lin2018structure,liu2025physical,jiang2025metallic,zhang2025crystal,hu2026observation,sun2026emergent,li2026pressure} naturally provide such
a mechanism: spin-singlet pairs formed between the two valleys inherit
their momentum separation $\mathbf Q$, yielding
$\Delta_{\mathbf R}
=\Delta_\eta e^{i\mathbf Q\cdot\mathbf R}
=\Delta_\eta(-1)^{R_x+R_y}$.  Unlike conventional pair-density-wave states, where translational symmetry breaking is accompanied by amplitude modulation,  the order parameter here changes sign between neighboring unit
cells while retaining a uniform magnitude. It breaks primitive translation and gauge symmetry separately but
preserves their combined operation, realizing a projective translation
symmetry without modulating gauge-invariant local observables. Notably, we show that the projective crystalline symmetry of the staggered condensate leaves a striking macroscopic imprint on Josephson transport: symmetry-enforced interference suppresses charge-$2e$ tunneling into a uniform superconductor, promoting coherent two-pair transfer and converting the staggered phase structure into a charge-$4e$-like Josephson response.

\begin{figure}[h!]
	\centering
	\includegraphics[width=1\linewidth]{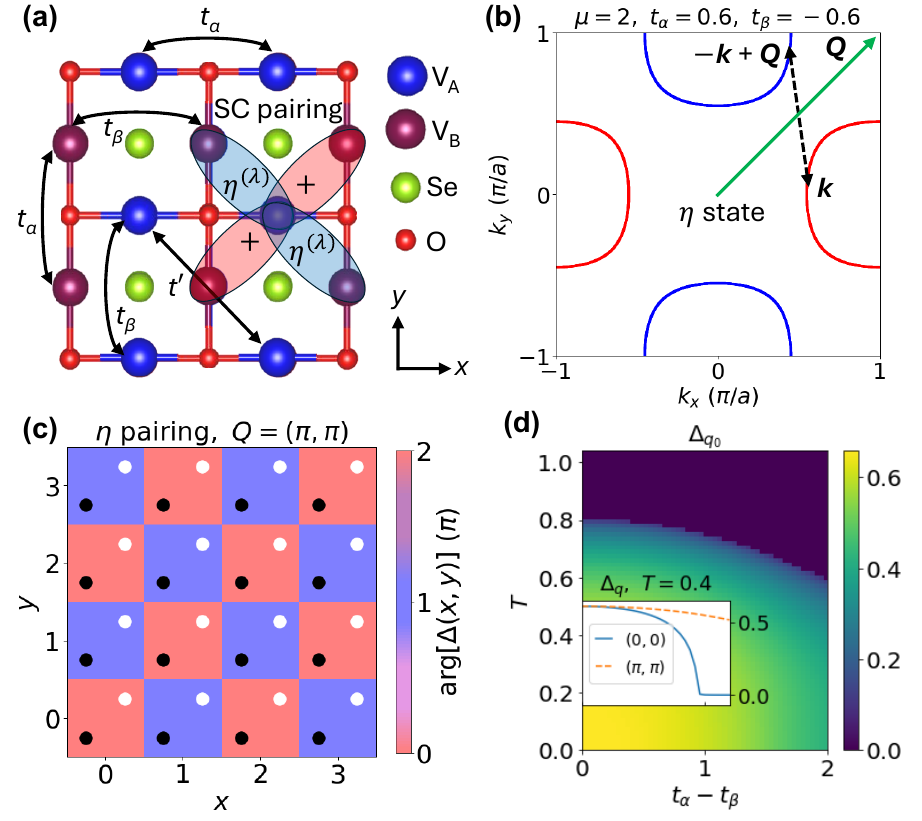}
	\caption{(a) Schematic of the hopping processes and intersublattice spin-singlet pairing considered in the model, including extended-$s$- and $d$-wave channels.
	(b) Representative Fermi surface exhibiting strong altermagnetic spin splitting and oppositely spin-polarized valleys at $X$ and $Y$.
	(c) Real-space structure of the $\eta$-like condensate, whose phase alternates by $\pi$ between neighboring unit cells while its magnitude remains uniform.
	(d) $\eta$-like-pairing amplitude as a function of temperature $T$ and hopping anisotropy $t_{\alpha}-t_{\beta}$, which controls the altermagnetic spin splitting. The inset compares the pairing amplitudes of uniform and $\eta$-like pairing at $T=0.4$.}
	\label{fig:fig1}
\end{figure}

{\sl Model.—--}Based on low-energy electronic structures of the 
altermagnetic configuration proposed for V$_2$Se$_2$O-family
materials~\cite{lin2018structure,liu2025physical,jiang2025metallic,zhang2025crystal}, we retain the V $d_{xy}$ orbitals on the two oppositely
polarized magnetic sublattices $A$ and $B$ within a single V$_2$O
layer. The large exchange splitting locks the low-energy spin
polarization to the sublattice, leaving the states
$|A\uparrow\rangle$ and $|B\downarrow\rangle$ near the Fermi level. The normal-state tight-binding Hamiltonian [illustrated in Fig.~\ref{fig:fig1}(a)] reads~\cite{cheng2026realistic}
\begin{align}
   &H=
\sum_{\mathbf R}\big(
t_{\alpha}c_{\mathbf R,A}^{\dagger}c_{\mathbf R+\hat{x},A}
+t_{\beta}c_{\mathbf R,A}^{\dagger}c_{\mathbf R+\hat{y},A}
+t_{\beta}c_{\mathbf R,B}^{\dagger}c_{\mathbf R+\hat{x},B}
\nonumber\\
&
+t_{\alpha}c_{\mathbf R,B}^{\dagger}c_{\mathbf R+\hat{y},B}
+t'\!\sum_{\tau,\boldsymbol{\delta}'}
c_{\mathbf R,\tau}^{\dagger}
c_{\mathbf R+\boldsymbol{\delta}',\tau}
+\mathrm{h.c.}\big)-\mu{N},
\label{eq:realspace_normal}
\end{align}
Here, $\tau=A,B$ labels the two oppositely spin-polarized magnetic sublattices, and $\boldsymbol{\delta}'=\hat{x}+\hat{y},\hat{x}-\hat{y}$ denotes the two inequivalent diagonal bonds, with the opposite directions included through the Hermitian conjugate. The directional hopping anisotropy is reversed between the two sublattices: the $A$ sector has hopping amplitudes $(t_{\alpha},t_{\beta})$ along the $(x,y)$ directions, whereas the $B$ sector has $(t_{\beta},t_{\alpha})$. Consequently, a $C_{4z}$ rotation, simultaneously maps the $A$ and $B$ sectors onto each other. Since the two sectors carry opposite spin polarizations, this realizes the spin-group symmetry $[C_2\Vert C_{4z}]$, with the accompanying $\pi$ spin rotation $C_2$ reversing the spin polarization. The sublattice-dependent hopping anisotropy therefore generates an altermagnetic spin splitting proportional to $t_{\alpha}-t_{\beta}$, which reverses sign under $x\leftrightarrow y$ and hence has the characteristic $d$-wave form. Finite $t_{\alpha}+t_{\beta}$ controls the relative matching between the two valleys at $X=(\pi,0)$ and $Y=(0,\pi)$, with perfect matching achieved at $t_{\alpha}+t_{\beta}=0$.

Since the low-energy states on the $A$ and $B$ sublattices are oppositely spin polarized, spin-singlet pairing necessarily occurs between the two magnetic sublattices. Projecting onto the low-energy states $c_{\mathbf R,A}\equiv c_{\mathbf R,A\uparrow}$ and $c_{\mathbf R,B}\equiv c_{\mathbf R,B\downarrow}$, we introduce an attractive pairing interaction
\begin{equation}
H_{\rm pair}
=
-\sum_{\mathbf R,\boldsymbol{\delta}}
V_{\boldsymbol{\delta}}\,
c_{\mathbf R,A}^{\dagger}
c_{\mathbf R+\boldsymbol{\delta},B}^{\dagger}
c_{\mathbf R+\boldsymbol{\delta},B}
c_{\mathbf R,A},
\label{eq:pairing}
\end{equation}
where $\boldsymbol{\delta}$ specifies the relative displacement between the paired electrons. The corresponding SC order parameter then reads
\begin{equation}
\Delta_{\boldsymbol{\delta}}(\mathbf R_c)
=
V_{\boldsymbol{\delta}}
\left\langle
c_{\mathbf R,A}
c_{\mathbf R+\boldsymbol{\delta},B}
\right\rangle,
\qquad
\mathbf R_c=\mathbf R+\frac{\boldsymbol{\delta}}{2},
\label{eq:bond_gap}
\end{equation}
where $\mathbf R_c$ is the center-of-mass coordinate of Cooper pair.

The internal symmetry of the Cooper pair is determined by the
transformation of the bond order parameter under the crystal
symmetries. For the two diagonal intersublattice bonds
\begin{equation}
\boldsymbol{\delta}_{+}
=
\frac{\hat{x}+\hat{y}}{2},
\qquad
\boldsymbol{\delta}_{-}
=
\frac{\hat{x}-\hat{y}}{2},
\label{eq:pair_bonds}
\end{equation}
a $C_{4z}$ rotation interchanges the two bond orientations,
$\boldsymbol{\delta}_{+}\leftrightarrow
\boldsymbol{\delta}_{-}$, up to reversal of the bond direction.
The two corresponding pairing channels are distinguished by their
relative phase on these symmetry-related bonds [as shown in Fig.~\ref{fig:fig1}(a)], yielding 
\begin{equation}
\Delta_{\boldsymbol{\delta}}(\mathbf R_c)
=
\eta_{\boldsymbol{\delta}}^{(\lambda)}
\Delta(\mathbf R_c),
\qquad
\lambda=s,d,
\label{eq:pair_symmetry}
\end{equation}
with  
$\eta_{\boldsymbol{\delta}_{+}}^{(s)}
=
\eta_{\boldsymbol{\delta}_{-}}^{(s)}
=1$ and $\eta_{\boldsymbol{\delta}_{+}}^{(d)}
=
-\eta_{\boldsymbol{\delta}_{-}}^{(d)}
=1$. The equal-sign combination is invariant under $C_{4z}$ and
corresponds to an extended-$s$-wave pair~\cite{reid2012d}, whereas the opposite-sign
combination changes sign under $C_{4z}$ and corresponds to a
$d$-wave pair. Both states are nonlocal bond-centered Cooper pairs;
their distinction lies solely in the transformation of their
relative-coordinate wave functions under the crystal symmetry.  For equal pairing strengths on the two symmetry-related bonds, the condensation energy depends only on $|\Delta_{\boldsymbol{\delta}_{+}}|^2+|\Delta_{\boldsymbol{\delta}_{-}}|^2$ and is insensitive to their relative sign. The extended-$s$- and $d$-wave channels are therefore equivalent at this level, allowing us to focus below on the center-of-mass structure of the condensate.

To describe a condensate with finite center-of-mass momentum $\mathbf q$, we write
$\Delta(\mathbf R_c)=|\Delta_{\mathbf q}|e^{i\mathbf q\cdot\mathbf R_c}$, such that
\begin{equation}
\Delta_{\boldsymbol{\delta}}(\mathbf R_c)
=
\eta_{\boldsymbol{\delta}}^{(\lambda)}
|\Delta_{\mathbf q}|
e^{i\mathbf q\cdot\mathbf R_c}.
\label{eq:full_gap}
\end{equation} 
The internal pairing symmetry
enters solely through the bond form factor
$\eta_{\boldsymbol{\delta}}^{(\lambda)}$, whereas the center-of-mass
structure is encoded by the phase factor
$e^{i\mathbf q\cdot\mathbf R_c}$. Since the bond form factor
$\eta_{\boldsymbol{\delta}}^{(\lambda)}=\pm1$ enters only through
the phase of the pairing amplitude, it drops out of the
quasiparticle spectrum and hence of the thermodynamic potential.
The extended-$s$- and $d$-wave channels therefore yield the same
free-energy functional for the center-of-mass condensate within the
present model and will not be distinguished hereafter. Applying mean-field treatment gives the Bogoliubov quasiparticle
energies $E_\nu(|\Delta_{\mathbf q}|^2,\mathbf q)$, and then, the corresponding
thermodynamic potential is
\begin{equation}
\Omega(|\Delta_{\mathbf q}|,\mathbf q)
=
\frac{N|\Delta_{\mathbf q}|^2}{V_{\lambda}}
- \frac{1}{\beta} \sum_{\nu} \ln\Big[ 2\cosh\Big(\frac{\beta E_{\nu}}{2}\Big) \Big], \label{eq:free_energy_realspace} 
\end{equation} 
where $\beta=1/(k_{\rm B}T)$. The equilibrium pairing amplitude and center-of-mass modulation are determined variationally from \begin{equation} \left(|\Delta_{\mathbf q_0}|,\mathbf q_0\right) = \underset{|\Delta_{\mathbf q}|,\mathbf q} {\operatorname{arg\,min}}\, \Omega(|\Delta_{\mathbf q}|,\mathbf q). \label{eq:minimization} \end{equation}

{\sl Numerical Results.---}The low-energy states, as shown in Fig.~\ref{fig:fig1}(b), form oppositely spin-polarized Fermi pockets centered at the $X=(\pi,0)$ and $Y=(0,\pi)$ valleys in V$_2$Se$_2$O-family altermagnets, which are separated by $\mathbf Q=(\pi,\pi)$. Pairing between these symmetry-related valleys directly favors the $\eta$-like condensate illustrated in Fig.~\ref{fig:fig1}(c), whose order parameter changes sign between neighboring unit cells while maintaining a uniform magnitude. Increasing altermagnetic spin splitting $|t_{\alpha}-t_{\beta}|$, as shown in the inset of Fig.~\ref{fig:fig1}(d), strongly suppresses uniform pairing as expected, whereas the $\eta$-like-pairing remains robust over a broad parameter regime.

The underlying origin becomes transparent under the staggered gauge transformation, $c_{\mathbf R B}=(-1)^{R_x+R_y}\widetilde c_{\mathbf R B}$, which absorbs the staggered phase of the $\eta$-like condensate into the $B$-sublattice fermion operator. The $\eta$-like-pairing problem is then mapped onto a uniform pairing problem  between the $A$ and $\widetilde B$ sectors. Under this transformation, hopping between opposite sublattices of the checkerboard parity changes sign for an odd displacement, whereas it remains unchanged for an even displacement, yielding the hopping structures,
\begin{equation}
A:(t_{\alpha},t_{\beta}),\quad B:(t_{\beta},t_{\alpha})
\longrightarrow
A:(t_{\alpha},t_{\beta}),\quad \widetilde B:(-t_{\beta},-t_{\alpha}),
\nonumber
\end{equation}
while the diagonal hopping $t'$ remains unchanged.  For $t_{\alpha}+t_{\beta}=0$,  the transformed hopping structure of the $\widetilde B$ sector, $(-t_{\beta},-t_{\alpha})$, becomes identical to that of the $A$ sector, $(t_{\alpha},t_{\beta})$. The two pairing are thus exactly matched irrespective of the magnitude of $|t_{\alpha}-t_{\beta}|$, i.e., increasing $|t_{\alpha}-t_{\beta}|$ does not introduce an energy mismatch between the two partners of a Cooper pair and hence does not act as a pair-breaking field. This distinction can also be seen explicitly from the quasiparticle spectra. In the limit $t_{\alpha}+t_{\beta}=0$, the normal-state Hamiltonian takes the form 
$H_{\mathbf k}=f_{0\mathbf k}\sigma_0+f_{z\mathbf k}\sigma_z$, 
where 
$f_{0\mathbf k}=-\mu+4t'\cos k_x\cos k_y$ and 
$f_{z\mathbf k}=2t_{\alpha}(\cos k_x-\cos k_y)$ 
satisfying
$f_{0,\mathbf k+\mathbf Q}=f_{0\mathbf k}$ and
$f_{z,\mathbf k+\mathbf Q}=-f_{z\mathbf k}$.
For uniform pairing, the positive-energy quasiparticle branches are 
$E_{\pm}^{\rm uniform}
=
\sqrt{f_{0\mathbf k}^{\,2}+\Delta^2g_{\mathbf k}^{\,2}}
\pm f_{z\mathbf k}$. Here, the altermagnetic splitting $f_{z\mathbf k}$ acts as a momentum-dependent effective Zeeman field~\cite{Zhang2024,whm1-76jc,6wxh-p4mc,PhysRevResearch.5.043171,rqp1-jtcb,mukasa2025finite, 3k12-2467, cvq6-b1nf}, generating an energy mismatch between the two pairing partners and thereby suppressing uniform spin-singlet superconductivity~\cite{mazin2025notes,chakraborty2025constraints,rqp1-jtcb}. In contrast, for spin-singlet $\eta$-like pairing, 
$E_{\pm}^{\eta}
=
\sqrt{
\left(f_{0\mathbf k}\pm f_{z\mathbf k}\right)^2
+\Delta^2g_{\mathbf k}^{\,2}
}$.  The altermagnetic splitting now enters the normal-state dispersion inside the square root, and therefore reshapes the quasiparticle dispersion without directly competing with the $\eta$-like-pairing gap, microscopically explaining the robustness of the $\eta$-like condensate against increasing altermagnetic splitting $|t_{\alpha}-t_{\beta}|$.

\begin{figure}[t]
	\centering
	\includegraphics[width=1\linewidth]{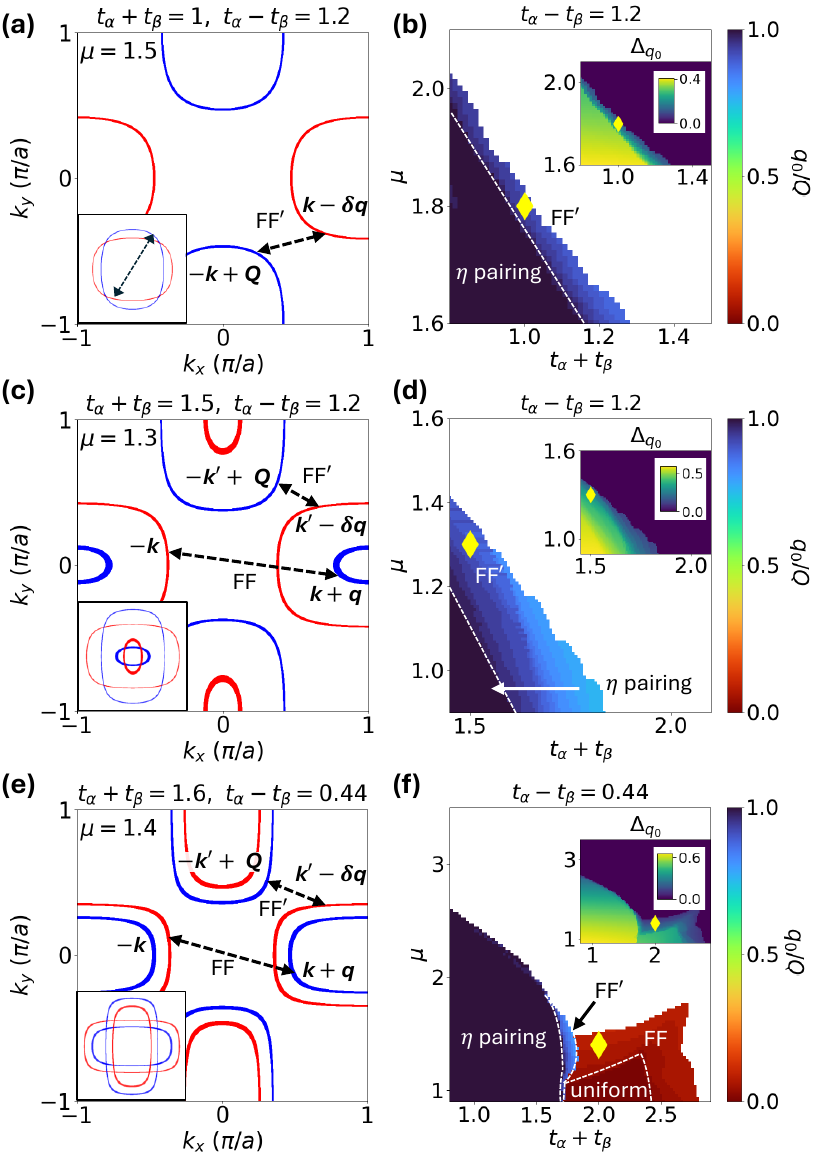}
	\caption{(a) Fermi surface with single pockets and finite valley mismatch, where the $\eta$-like state evolves into an intervalley FF$'$ state with pairing momentum $\mathbf{Q}-\delta\mathbf{q}$ (i.e.,  an FF-like modulation of the staggered $\eta$-like-pairing state). The inset shows the corresponding small momentum shift $\delta\mathbf{q}$ in the folded Brillouin zone.  (b) Pairing momentum and SC gap (inset) for strong altermagnetic splitting, showing the $\eta$-like-to-FF$'$ transition occurring at large valley mismatch. (c) Fermi surface with two overlapping pockets, allowing finite-momentum intravalley pairing. The inset shows the folded Brillouin zone.  
(d) Equilibrium Cooper-pair momentum and SC gap (inset) as functions of $t_{\alpha}+t_{\beta}$ and $\mu$. (e) Fermi surface with double pockets, enabling an intravalley FF state with pairing momentum $\mathbf{q}_0$. The inset shows the folded Brillouin zone.  
(f) Pairing momentum and SC gap for weak altermagnetic splitting, showing the competition among the $\eta$-like, FF$'$, FF, and uniform-phase SC states. We set $t'=-1$ throughout.}
	\label{fig:fig2}
\end{figure}

The stability of $\eta$-like pairing is controlled by the mismatch between the two valleys, governed in our model by $t_{\alpha}+t_{\beta}$. For the single-pocket electronic structure in Fig.~\ref{fig:fig2}(a), increasing the valley mismatch progressively frustrates intervalley pairing, leading to the evolution from the commensurate $\eta$-like state to an incommensurate state and eventually to the normal state, as demonstrated in Figs.~\ref{fig:fig2}(b). In the intermediate state, the pairing momentum remains close to $\mathbf Q$ but continuously deviates from it with increasing valley mismatch. We denote this state as FF$'$ to emphasize its continuity from $\eta$-like pairing, which can be viewed as an FF-like modulation~\cite{Zhang2024,whm1-76jc,6wxh-p4mc,PhysRevResearch.5.043171,rqp1-jtcb} of the staggered $\eta$-like-pairing state. The emergence of FF$'$ suggests that intervalley pairing can accommodate a finite valley mismatch by continuously adjusting its center-of-mass momentum while retaining its staggered character, as expected from the physics of FFLO-like pairing. 

The Fermi-surface topology varies across the V$_2$Se$_2$O family, with some members exhibiting multiple pockets beyond the single-pocket regime considered above. While recent experiments reveal single spin-split pockets around the $X$ and $Y$ points in KV$_2$Se$_2$O~\cite{Jiang2025,bf1n-sxdl,r8nc-dpt8,yang2025altermagnetic,yan2025sdw,yang2026emergent}, double pockets have been observed in RbV$_{1-\delta}$Te$_2$O~\cite{zhang2025crystal,yin2026discovery}. Typically, the additional pockets remain small, as shown in Fig.~\ref{fig:fig2}(c), and therefore do not destabilize $\eta$-like pairing, which persists over broad ranges of chemical potential $\mu$ and valley mismatch $t_{\alpha}+t_{\beta}$ [Fig.~\ref{fig:fig2}(d)]. As in the single-pocket case, increasing the valley mismatch in this regime drives the $\eta$-like state into FF$'$ before superconductivity eventually gives way to the normal state. Only when the additional pockets become sufficiently large [as shown in Fig.~\ref{fig:fig2}(e)] do they qualitatively reshape this phase evolution by rendering intravalley pairing competitive. The system can then evolve from the $\eta$-like state through FF$'$ into intravalley uniform or FF pairing before superconductivity is ultimately destroyed, as shown in Fig.~\ref{fig:fig2}(f). Thus, in this regime, the loss of $\eta$-like pairing no longer coincides with the loss of superconductivity, but marks a transition from intervalley to intravalley pairing. In particular, the transition from FF$'$ to either FF or uniform pairing is first order, reflecting an abrupt reconstruction of the condensate from intervalley to intravalley pairing.

\begin{figure}[h!]
	\centering
	\includegraphics[width=1\linewidth]{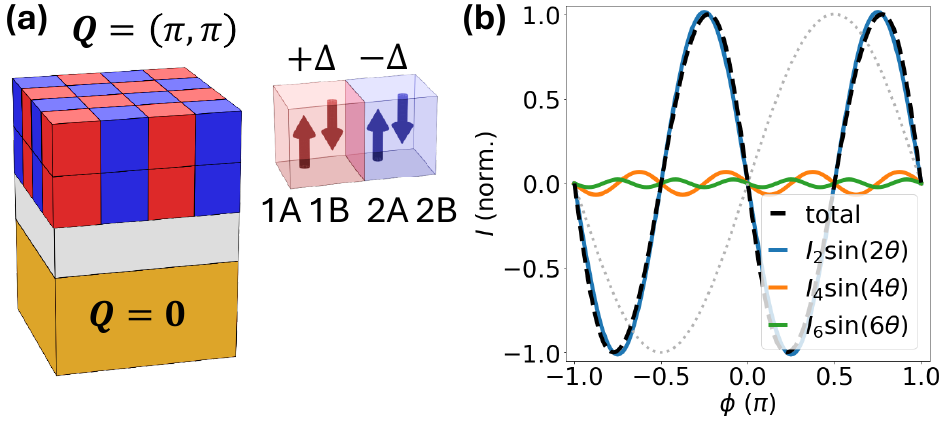}
    \caption{(a) Schematic of a Josephson junction between a staggered $\eta$-like-pairing SC and a uniform SC. The staggered pairing doubles the unit cell, which contains two primitive cells and four sublattices.
    (b) Current-phase relation (CPR) and its harmonic decomposition. All odd harmonics vanish due to the staggered structure of the $\eta$-like condensate, leaving only even-harmonic contributions. The CPR is dominated by the second harmonic $\sin(2\theta)$, with smaller $\sin(4\theta)$ and $\sin(6\theta)$ components. In the simulations, $\mu=2$, $t'=-1$, and $t_z=1$, with $t_{\alpha}=-t_{\beta}=0.6$ for the $\eta$-like-pairing SC.}
	\label{fig:fig3}
\end{figure}

These results establish $\eta$-like pairing as a robust SC  state in V$_2$Se$_2$O-family altermagnets, emerging over a broad region of the SC phase diagram.  Having established this robustness, we now turn to its experimental identification.  A defining feature that distinguishes the $\eta$-like state from conventional superconductivity is its staggered phase structure, while conventional quasiparticle spectroscopies such as ARPES and STM, which predominantly probe the quasiparticle excitation spectrum, have difficulty resolving this phase structure. A phase-sensitive probe is therefore essential for unambiguously identifying the $\eta$-like state. To this end, we propose a Josephson junction between an $\eta$-like-paired SC state and a conventional uniform-phase SC electrode, such as Nb or NbN. We calculate the current-phase relation (CPR) using the tight-binding model for the V$_2$Se$_2$O-family altermagnets introduced in Eq.~(\ref{eq:realspace_normal}). The junction consists of an $\eta$-like-paired SC state, a nonmagnetic metallic barrier, and a uniform-phase SC electrode, as illustrated in Fig.~\ref{fig:fig3}(a). As shown in Fig.~\ref{fig:fig3}(b), the calculated CPR contains exclusively even harmonics,
\begin{equation}
I(\phi)=I_{2}\sin(2\phi)+I_{4}\sin(4\phi)+I_{6}\sin(6\phi)+\cdots,
\end{equation}
with the second harmonic $I_2\sin(2\phi)$ providing the dominant contribution. The absence of odd harmonics originates from destructive interference associated with the staggered phase structure of the $\eta$-like condensate. For a symmetry-balanced vertical interface, neighboring unit cells contribute equally to Josephson tunneling but carry opposite signs of the $\eta$-like order parameter,
$\Delta_{\eta,2}=-\Delta_{\eta,1}$. Thus, the $n$-Cooper-pair tunneling amplitudes from two neighboring cells satisfy
\begin{equation}
\mathcal{J}_{n}^{(1)}+\mathcal{J}_{n}^{(2)}
\propto
\Delta_{\eta,1}^{\,n}+\Delta_{\eta,2}^{\,n}
=
\left[1+(-1)^n\right]\Delta_{\eta,1}^{\,n},
\label{eq:josephson_interference}
\end{equation}
which vanishes for odd $n$ but remains finite for even $n$. Thus, all odd Josephson harmonics cancel, while even harmonics survive. The leading surviving process corresponds to coherent tunneling of two Cooper pairs and produces the dominant $I_2\sin(2\phi)$ component. This gives rise to a charge-$4e$-like Josephson response even though the bulk $\eta$-like condensate itself consists of charge-$2e$ Cooper pairs.  We also find that the CPR is strongly suppressed upon entering the FF$'$, where the incommensurate
Cooper-pair momentum introduces spatially varying phase differences
across the junction, leading to destructive interference
among different tunneling contributions.

{\sl Discussion.---}The broken symmetry of the $\eta$-like condensate resides in its staggered phase structure, whereby primitive translation reverses the sign of the order parameter while leaving its local magnitude uniform, making it difficult to identify via purely local probes. Josephson interference provides a natural way to expose this hidden structure as it directly probes the SC phase. The survival of higher-order pair transfer, with the suppression of leading-order tunneling at a symmetry-balanced interface, converts the microscopic sign alternation of a charge-$2e$ condensate into a charge-$4e$-like Josephson response, providing a distinct phase-sensitive signature when local observables appear
featureless. While interface imperfections, local disorder, and asymmetric tunneling in realistic junctions may locally lift the cancellation of the leading charge-$2e$ Josephson coupling, the staggered phase structure still favors an enhanced relative contribution from higher-order charge-$4e$ processes.

While symmetry-related valleys at distinct high-symmetry momenta are common in multivalley materials, these special momenta often allow intravalley pairing to remain equivalent to conventional zero-momentum pairing modulo a reciprocal lattice vector, such that the valley structure itself does not necessarily impart a nontrivial center-of-mass momentum to the condensate. The essential ingredient is a symmetry-related pair of valleys carrying opposite spin polarization and a nonzero commensurate momentum separation. V$_2$Se$_2$O-family altermagnets represent a distinct route to finite-momentum superconductivity, in which magnetic crystal symmetry relates valleys at distinct momenta while simultaneously enforcing their opposite-spin character, converting a symmetry-defined momentum structure into a nontrivial center-of-mass momentum for spin-singlet Cooper pairs. Distinct spin-singlet channels, such as extended $s$-wave and $d$-wave pairing, acquire the same symmetry-selected center-of-mass momentum and realize the same staggered superconductivity, demonstrating that the emergence of $\eta$-like pairing is governed primarily by the crystal-symmetry relation between the low-energy sectors rather than by the internal symmetry of the Cooper pair. Moreover, the $\eta$-like-pairing mechanism is essentially independent of the interlayer magnetic order. The N\'eel orders of adjacent layers may be aligned either parallel (C-type) or antiparallel (G-type) \cite{yang2025observation, sun2025antiferromagnetic,gu2026real}, but this distinction only changes the relative magnetic orientation between neighboring layers. The electronic structure of V$_2$Se$_2$O-family compounds is highly two-dimensional, with negligible interlayer hopping and predominantly intralayer SC pairing. Thus, interlayer magnetic stacking is not expected to affect the pairing or the symmetry-selected staggered structure of the condensate.  Multivalley altermagnets thus provide a natural materials platform for symmetry-enforced finite-momentum superconductivity, in which crystalline magnetism determines not only the spin structure of the normal state but also the translational structure of the SC condensate, opening new opportunities for engineering unconventional SC states and their functionalities in SC electronics and quantum technologies.

\appendix

{\it Acknowledgments.---}This work is supported by the Hong Kong Research Grants Council (CRS$\_$HKUST603/25, C6046-24G, 16306722, 16304523, 16311125, and 16311226).
\bibliography{apssamp}% Produces the bibliography via BibTeX.

@article{feng2022anomalous,
  author = {Feng, Z. and Zhou, X. and {\v S}mejkal, L. and Wu, L. and Zhu, Z. and Guo, H. and Gonz{\'a}lez-Hern{\'a}ndez, R. and Wang, X. and Yan, H. and Qin, P. and Zhang, X. and Wu, H. and Chen, H. and Meng, Z. and Liu, L. and Xia, Z. and Sinova, J. and Jungwirth, T. and Liu, Z.},
  title = {An anomalous {H}all effect in altermagnetic ruthenium dioxide},
  journal = {Nat. Electron.},
  volume = {5},
  number = {11},
  pages = {735--743},
  year = {2022},
  doi = {10.1038/s41928-022-00866-z}
}

@article{liao2024separation,
  title = {Separation of Inverse Altermagnetic Spin-Splitting Effect from Inverse Spin Hall Effect in {${\mathrm{RuO}}_{2}$}},
  author = {Liao, Ching-Te and Wang, Yu-Chun and Tien, Yu-Cheng and Huang, Ssu-Yen and Qu, Danru},
  journal = {Phys. Rev. Lett.},
  volume = {133},
  issue = {5},
  pages = {056701},
  numpages = {5},
  year = {2024},
  month = {Jul},
  publisher = {American Physical Society},
  doi = {10.1103/PhysRevLett.133.056701},
  url = {https://link.aps.org/doi/10.1103/PhysRevLett.133.056701}
}

@article{sato2024altermagnetic,
  title = {Altermagnetic Anomalous Hall Effect Emerging from Electronic Correlations},
  author = {Sato, Toshihiro and Haddad, Sonia and Fulga, Ion Cosma and Assaad, Fakher F. and van den Brink, Jeroen},
  journal = {Phys. Rev. Lett.},
  volume = {133},
  issue = {8},
  pages = {086503},
  numpages = {6},
  year = {2024},
  month = {Aug},
  publisher = {American Physical Society},
  doi = {10.1103/PhysRevLett.133.086503},
  url = {https://link.aps.org/doi/10.1103/PhysRevLett.133.086503}
}

@article{ma2021multifunctional,
  author = {Ma, H.-Y. and Hu, M. and Li, N. and Liu, J. and Yao, W. and Jia, J.-F. and Liu, J.},
  title = {Multifunctional antiferromagnetic materials with giant piezomagnetism and noncollinear spin current},
  journal = {Nat. Commun.},
  volume = {12},
  number = {1},
  pages = {2846},
  year = {2021},
  doi = {10.1038/s41467-021-23127-7}
}

@article{fu2025all,
  author = {Fu, P.-H. and Lv, Q. and Xu, Y. and Cayao, J. and Liu, J.-F. and Yu, X.-L.},
  title = {All-electrically controlled spintronics in altermagnetic heterostructures},
  journal = {npj Quantum Mater.},
  volume = {10},
  number = {1},
  pages = {111},
  year = {2025},
  doi = {10.1038/s41535-025-00827-7}
}

@article{rqp1-jtcb,
  title = {Altermagnetism-induced noncollinear superconducting diode effect and unidirectional superconducting transport},
  author = {Yang, F. and Chen, L. Q.},
  journal = {Phys. Rev. B},
  volume = {112},
  issue = {22},
  pages = {L220502},
  numpages = {8},
  year = {2025},
  month = {Dec},
  publisher = {American Physical Society},
  doi = {10.1103/rqp1-jtcb},
  url = {https://link.aps.org/doi/10.1103/rqp1-jtcb}
}

@article{fukaya2025superconducting,
author = {Y. Fukaya and B. Lu and K. Yada and Y. Tanaka and J. Cayao},
title = {Superconducting phenomena in systems with unconventional magnets},
journal = {J. Phys.: Condens. Matter},
year = {2025},
volume = {37},
number = {31},
pages = {313003},
doi = {10.1088/1361-648X/adf1cf}
}

@article{sun2025spin,
  title = {Spin relaxation and transport behavior in $d$-wave altermagnetic systems},
  author = {Sun, Y. J. and Yang, F. and Chen, L. Q.},
  journal = {Phys. Rev. B},
  volume = {112},
  issue = {2},
  pages = {024412},
  numpages = {12},
  year = {2025},
  month = {Jul},
  publisher = {American Physical Society},
  doi = {10.1103/v12v-gl4n},
  url = {https://link.aps.org/doi/10.1103/v12v-gl4n}
}

@article{fukaya2025josephson,
  title = {Josephson effect and odd-frequency pairing in superconducting junctions with unconventional magnets},
  author = {Fukaya, Yuri and Maeda, Kazuki and Yada, Keiji and Cayao, Jorge and Tanaka, Yukio and Lu, Bo},
  journal = {Phys. Rev. B},
  volume = {111},
  issue = {6},
  pages = {064502},
  numpages = {15},
  year = {2025},
  month = {Feb},
  publisher = {American Physical Society},
  doi = {10.1103/PhysRevB.111.064502},
  url = {https://link.aps.org/doi/10.1103/PhysRevB.111.064502}
}

@article{PhysRevB.111.184515,
  title = {Orientation-dependent transport in junctions formed by $d$-wave altermagnets and $d$-wave superconductors},
  author = {Zhao, Wenjun and Fukaya, Yuri and Burset, Pablo and Cayao, Jorge and Tanaka, Yukio and Lu, Bo},
  journal = {Phys. Rev. B},
  volume = {111},
  issue = {18},
  pages = {184515},
  numpages = {11},
  year = {2025},
  month = {May},
  publisher = {American Physical Society},
  doi = {10.1103/PhysRevB.111.184515},
  url = {https://link.aps.org/doi/10.1103/PhysRevB.111.184515}
}

@article{PhysRevLett.131.076003,
  title = {{DC} {J}osephson Effect in Altermagnets},
  author = {Ouassou, Jabir Ali and Brataas, Arne and Linder, Jacob},
  journal = {Phys. Rev. Lett.},
  volume = {131},
  issue = {7},
  pages = {076003},
  numpages = {6},
  year = {2023},
  month = {Aug},
  publisher = {American Physical Society},
  doi = {10.1103/PhysRevLett.131.076003},
  url = {https://link.aps.org/doi/10.1103/PhysRevLett.131.076003}
}

@article{PhysRevB.108.054511,
  title = {Andreev reflection in altermagnets},
  author = {Sun, Chi and Brataas, Arne and Linder, Jacob},
  journal = {Phys. Rev. B},
  volume = {108},
  issue = {5},
  pages = {054511},
  numpages = {14},
  year = {2023},
  month = {Aug},
  publisher = {American Physical Society},
  doi = {10.1103/PhysRevB.108.054511},
  url = {https://link.aps.org/doi/10.1103/PhysRevB.108.054511}
}

@article{dlpb-gfct,
  title = {Intra-Unit-Cell Singlet Pairing Mediated by Altermagnetic Fluctuations},
  author = {Wu, Yi-Ming and Wang, Yuxuan and Fernandes, Rafael M.},
  journal = {Phys. Rev. Lett.},
  volume = {135},
  issue = {15},
  pages = {156001},
  numpages = {8},
  year = {2025},
  month = {Oct},
  publisher = {American Physical Society},
  doi = {10.1103/dlpb-gfct},
  url = {https://link.aps.org/doi/10.1103/dlpb-gfct}
}

@article{PhysRevX.12.031042,
  title = {Beyond Conventional Ferromagnetism and Antiferromagnetism: A Phase with Nonrelativistic Spin and Crystal Rotation Symmetry},
  author = {{\ifmmode \check{S}\else \v{S}\fi{}}mejkal, Libor and Sinova, Jairo and Jungwirth, Tomas},
  journal = {Phys. Rev. X},
  volume = {12},
  issue = {3},
  pages = {031042},
  numpages = {16},
  year = {2022},
  month = {Sep},
  publisher = {American Physical Society},
  doi = {10.1103/PhysRevX.12.031042},
  url = {https://link.aps.org/doi/10.1103/PhysRevX.12.031042}
}

@article{PhysRevX.12.040501,
  title = {Emerging Research Landscape of Altermagnetism},
  author = {{\ifmmode \check{S}\else \v{S}\fi{}}mejkal, Libor and Sinova, Jairo and Jungwirth, Tomas},
  journal = {Phys. Rev. X},
  volume = {12},
  issue = {4},
  pages = {040501},
  numpages = {27},
  year = {2022},
  month = {Dec},
  publisher = {American Physical Society},
  doi = {10.1103/PhysRevX.12.040501},
  url = {https://link.aps.org/doi/10.1103/PhysRevX.12.040501}
}

@article{Zhang2024,
  author  = {Zhang, Song-Bo and Hu, Lun-Hui and Neupert, Titus},
  title   = {{Finite-momentum Cooper pairing in proximitized altermagnets}},
  journal = {Nat. Commun.},
  volume  = {15},
  number  = {1},
  pages   = {1801},
  year    = {2024},
  doi     = {10.1038/s41467-024-45951-3}
}

@article{whm1-76jc,
  title = {{Unconventional superconductivity of an altermagnetic metal: Polarized BCS and inhomogeneous FFLO states}},
  author = {Hu, Hui and Liu, Zhao and Liu, Xia-Ji},
  journal = {Phys. Rev. B},
  volume = {112},
  issue = {18},
  pages = {184501},
  numpages = {11},
  year = {2025},
  month = {Nov},
  publisher = {American Physical Society},
  doi = {10.1103/whm1-76jc},
  url = {https://link.aps.org/doi/10.1103/whm1-76jc}
}

@article{6wxh-p4mc,
  title = {{Electron pairing by dispersive phonons in altermagnets: Reentrant superconductivity and continuous transition to finite momentum superconducting state}},
  author = {Iorsh, I. V.},
  journal = {Phys. Rev. B},
  volume = {111},
  issue = {22},
  pages = {L220503},
  numpages = {6},
  year = {2025},
  month = {Jun},
  publisher = {American Physical Society},
  doi = {10.1103/6wxh-p4mc},
  url = {https://link.aps.org/doi/10.1103/6wxh-p4mc}
}

@article{PhysRevResearch.5.043171,
  title = {{Fulde-Ferrell-Larkin-Ovchinnikov state induced by antiferromagnetic order in $\ensuremath{\kappa}$-type organic conductors}},
  author = {Sumita, Shuntaro and Naka, Makoto and Seo, Hitoshi},
  journal = {Phys. Rev. Res.},
  volume = {5},
  issue = {4},
  pages = {043171},
  numpages = {17},
  year = {2023},
  month = {Nov},
  publisher = {American Physical Society},
  doi = {10.1103/PhysRevResearch.5.043171},
  url = {https://link.aps.org/doi/10.1103/PhysRevResearch.5.043171}
}

@article{PhysRevLett.126.127701,
  title = {{Efficient Electrical Spin Splitter Based on Nonrelativistic Collinear Antiferromagnetism}},
  author = {Gonz{\'a}lez-Hern{\'a}ndez, Rafael and {\ifmmode \check{S}\else \v{S}\fi{}}mejkal, Libor and V{\'y}born{\'y}, Karel and Yahagi, Yuta and Sinova, Jairo and Jungwirth, Tom{\'a}{\ifmmode \check{s}\else \v{s}\fi{}} and {\ifmmode \check{Z}\else \v{Z}\fi{}}elezn{\'y}, Jakub},
  journal = {Phys. Rev. Lett.},
  volume = {126},
  issue = {12},
  pages = {127701},
  numpages = {6},
  year = {2021},
  month = {Mar},
  publisher = {American Physical Society},
  doi = {10.1103/PhysRevLett.126.127701},
  url = {https://link.aps.org/doi/10.1103/PhysRevLett.126.127701}
}

@article{PhysRevLett.130.216701,
  title = {{Efficient Spin-to-Charge Conversion via Altermagnetic Spin Splitting Effect in Antiferromagnet ${\mathrm{RuO}}_{2}$}},
  author = {Bai, H. and Zhang, Y. C. and Zhou, Y. J. and Chen, P. and Wan, C. H. and Han, L. and Zhu, W. X. and Liang, S. X. and Su, Y. C. and Han, X. F. and Pan, F. and Song, C.},
  journal = {Phys. Rev. Lett.},
  volume = {130},
  issue = {21},
  pages = {216701},
  numpages = {6},
  year = {2023},
  month = {May},
  publisher = {American Physical Society},
  doi = {10.1103/PhysRevLett.130.216701},
  url = {https://link.aps.org/doi/10.1103/PhysRevLett.130.216701}
}

@article{PhysRevB.111.054501,
  title = {{Unconventional $p$-wave and finite-momentum superconductivity induced by altermagnetism through the formation of Bogoliubov Fermi surface}},
  author = {Hong, SeungBeom and Park, Moon Jip and Kim, Kyoung-Min},
  journal = {Phys. Rev. B},
  volume = {111},
  issue = {5},
  pages = {054501},
  numpages = {9},
  year = {2025},
  month = {Feb},
  publisher = {American Physical Society},
  doi = {10.1103/PhysRevB.111.054501},
  url = {https://link.aps.org/doi/10.1103/PhysRevB.111.054501}
}

@article{Yang2017,
  author  = {Yang, F. and Wu, M. W.},
  title   = {{Gapped superconductivity with all symmetries in InSb (110) quantum wells in proximity to $s$-wave superconductor in Fulde-Ferrell-Larkin-Ovchinnikov phase or with a supercurrent}},
  journal = {Phys. Rev. B},
  volume  = {95},
  pages   = {075304},
  year    = {2017},
  doi     = {10.1103/PhysRevB.95.075304}
}

@article{52wh-1z5y,
  title = {Persistent Spin Currents in Superconducting Altermagnets},
  author = {Monkman, Kyle and Weng, Joan and Heinsdorf, Niclas and Nocera, Alberto and Barlas, Yafis and Franz, Marcel},
  journal = {Phys. Rev. X},
  volume = {16},
  issue = {1},
  pages = {011057},
  numpages = {21},
  year = {2026},
  month = {Mar},
  publisher = {American Physical Society},
  doi = {10.1103/52wh-1z5y},
  url = {https://link.aps.org/doi/10.1103/52wh-1z5y}
}

@article{PhysRevB.108.184505,
  title = {Topological superconductivity in two-dimensional altermagnetic metals},
  author = {Zhu, Di and Zhuang, Zheng-Yang and Wu, Zhigang and Yan, Zhongbo},
  journal = {Phys. Rev. B},
  volume = {108},
  issue = {18},
  pages = {184505},
  numpages = {13},
  year = {2023},
  month = {Nov},
  publisher = {American Physical Society},
  doi = {10.1103/PhysRevB.108.184505},
  url = {https://link.aps.org/doi/10.1103/PhysRevB.108.184505}
}

@article{7h2t-wmyr,
  title = {Proximity-induced superconductivity and emerging topological phases in altermagnet-based heterostructures},
  author = {Alam, Ohidul and Pal, Amartya and Dutta, Paramita and Saha, Arijit},
  journal = {Phys. Rev. B},
  volume = {113},
  issue = {15},
  pages = {155429},
  numpages = {17},
  year = {2026},
  month = {Apr},
  publisher = {American Physical Society},
  doi = {10.1103/7h2t-wmyr},
  url = {https://link.aps.org/doi/10.1103/7h2t-wmyr}
}

@article{PhysRevB.75.115103,
  title = {Fermi liquid instabilities in the spin channel},
  author = {Wu, Congjun and Sun, Kai and Fradkin, Eduardo and Zhang, Shou-Cheng},
  journal = {Phys. Rev. B},
  volume = {75},
  issue = {11},
  pages = {115103},
  numpages = {25},
  year = {2007},
  month = {Mar},
  publisher = {American Physical Society},
  doi = {10.1103/PhysRevB.75.115103},
  url = {https://link.aps.org/doi/10.1103/PhysRevB.75.115103}
}

@article{naka2019spin,
  title={Spin current generation in organic antiferromagnets},
  author={Naka, Makoto and Hayami, Satoru and Kusunose, Hiroaki and Yanagi, Yuki and Motome, Yukitoshi and Seo, Hitoshi},
  journal={Nat. Commun.},
  volume={10},
  number={1},
  pages={4305},
  year={2019},
  publisher={Nature Publishing Group UK London},
  doi={10.1038/s41467-019-12229-y},
  url={https://doi.org/10.1038/s41467-019-12229-y}
}

@article{smejkalCrystalTimereversalSymmetry2020,
  title={Crystal time-reversal symmetry breaking and spontaneous {H}all effect in collinear antiferromagnets},
  author={{\v{S}}mejkal, Libor and Gonz{\'a}lez-Hern{\'a}ndez, Rafael and Jungwirth, T. and Sinova, J.},
  journal={Sci. Adv.},
  volume={6},
  number={23},
  pages={eaaz8809},
  year={2020},
  publisher={American Association for the Advancement of Science},
  doi={10.1126/sciadv.aaz8809},
  url={https://doi.org/10.1126/sciadv.aaz8809}
}

@article{ahn2019antiferromagnetism,
  title={Antiferromagnetism in {R}u{O}\(_{2}\) as \(d\)-wave {P}omeranchuk instability},
  author={Ahn, Kyo-Hoon and Hariki, Atsushi and Lee, Kwan-Woo and Kune{\v{s}}, Jan},
  journal={Phys. Rev. B},
  volume={99},
  number={18},
  pages={184432},
  year={2019},
  publisher={APS},
  doi={10.1103/PhysRevB.99.184432},
  url={https://doi.org/10.1103/PhysRevB.99.184432}
}

@article{yuanGiantMomentumdependentSpin2020,
  title={Giant momentum-dependent spin splitting in centrosymmetric low-\({Z}\) antiferromagnets},
  author={Yuan, Lin-Ding and Wang, Zhi and Luo, Jun-Wei and Rashba, Emmanuel I. and Zunger, Alex},
  journal={Phys. Rev. B},
  volume={102},
  number={1},
  pages={014422},
  year={2020},
  publisher={APS},
  doi={10.1103/PhysRevB.102.014422},
  url={https://doi.org/10.1103/PhysRevB.102.014422}
}

@article{hayamiMomentumDependentSpinSplitting2019,
  title={Momentum-dependent spin splitting by collinear antiferromagnetic ordering},
  author={Hayami, Satoru and Yanagi, Yuki and Kusunose, Hiroaki},
  journal={J. Phys. Soc. Jpn.},
  volume={88},
  number={12},
  pages={123702},
  year={2019},
  publisher={The Physical Society of Japan},
  doi={10.7566/JPSJ.88.123702},
  url={https://doi.org/10.7566/JPSJ.88.123702}
}

@article{mazinPredictionUnconventionalMagnetism2021,
  title={Prediction of unconventional magnetism in doped {F}e{S}b\(_{2}\)},
  author={Mazin, Igor I. and Koepernik, Klaus and Johannes, Michelle D. and Gonz{\'a}lez-Hern{\'a}ndez, Rafael and {\v{S}}mejkal, Libor},
  journal={Proc. Natl. Acad. Sci. U.S.A.},
  volume={118},
  number={42},
  pages={e2108924118},
  year={2021},
  publisher={National Academy of Sciences},
  doi={10.1073/pnas.2108924118},
  url={https://doi.org/10.1073/pnas.2108924118}
}

@article{liuSpinGroupSymmetryMagnetic2022,
  title={Spin-group symmetry in magnetic materials with negligible spin-orbit coupling},
  author={Liu, Pengfei and Li, Jiayu and Han, Jingzhi and Wan, Xiangang and Liu, Qihang},
  journal={Phys. Rev. X},
  volume={12},
  number={2},
  pages={021016},
  year={2022},
  publisher={APS},
  doi={10.1103/PhysRevX.12.021016},
  url={https://doi.org/10.1103/PhysRevX.12.021016}
}

@article{cheng2026realistic,
  title={Realistic tight-binding model for {V}\(_{2}\){S}e\(_{2}\){O}-family altermagnets},
  author={Cheng, Xingkai and Gao, Yifan and Peng, Junjie and Liu, Junwei},
  journal={arXiv:2602.09465},
  year={2026},
  url={https://arxiv.org/abs/2602.09465}
}

@article{hu2026observation,
  title={Observation of spin-valley locked nodal lines in a quasi-2D altermagnet},
  author={Hu, Quanxin and Cheng, Xingkai and Duan, Qingchen and Hu, Yudong and Jiang, Bei and Xiao, Yusen and Li, Yaqi and Pan, Mojun and Deng, Liwei and Liu, Changchao and others},
  journal={arXiv:2601.02883},
  year={2026},
  url={https://arxiv.org/abs/2601.02883}
}

@article{mazin2025notes,
  title={Notes on altermagnetism and superconductivity},
  author={Mazin, Igor I},
  journal={AAPPS Bull.},
  volume={35},
  number={1},
  pages={18},
  year={2025},
  publisher={Springer},
  doi={10.1007/s43673-025-00158-6},
  url={https://doi.org/10.1007/s43673-025-00158-6}
}

@article{chakraborty2025constraints,
  title={Constraints on superconducting pairing in altermagnets},
  author={Chakraborty, Debmalya and Black-Schaffer, Annica M},
  journal={Phys. Rev. B},
  volume={112},
  number={1},
  pages={014516},
  year={2025},
  publisher={APS},
  doi = {10.1103/zylh-rqxl},
  url = {https://link.aps.org/doi/10.1103/zylh-rqxl}
}

@article{xu2025chemical,
  title={Chemical design of monolayer altermagnets},
  author={Xu, Runzhang and Gao, Yifan and Liu, Junwei},
  journal={Natl. Sci. Rev.},
  volume={12},
  number={5},
  pages={nwaf528},
  year={2025},
  publisher={Oxford University Press},
  doi={10.1093/nsr/nwaf528},
  url={https://doi.org/10.1093/nsr/nwaf528}
}

@article{lin2018structure,
  title = {Structure and physical properties of ${\mathrm{CsV}}_{2}{\mathrm{Se}}_{2\ensuremath{-}x}\mathrm{O}$ and ${\mathrm{V}}_{2}{\mathrm{Se}}_{2}\mathrm{O}$},
  author = {Lin, Hai and Si, Jin and Zhu, Xiyu and Cai, Kehan and Li, Hao and Kong, Lu and Yu, Xiaodong and Wen, Hai-Hu},
  journal = {Phys. Rev. B},
  volume = {98},
  issue = {7},
  pages = {075132},
  numpages = {5},
  year = {2018},
  month = {Aug},
  publisher = {American Physical Society},
  doi = {10.1103/PhysRevB.98.075132},
  url = {https://link.aps.org/doi/10.1103/PhysRevB.98.075132}
}

@article{hu2025catalog,
  title = {Catalog of $C$-Paired Spin-Momentum Locking in Antiferromagnetic Systems},
  author = {Hu, Mengli and Cheng, Xingkai and Huang, Zhenqiao and Liu, Junwei},
  journal = {Phys. Rev. X},
  volume = {15},
  issue = {2},
  pages = {021083},
  numpages = {11},
  year = {2025},
  month = {Jun},
  publisher = {American Physical Society},
  doi = {10.1103/PhysRevX.15.021083},
  url = {https://link.aps.org/doi/10.1103/PhysRevX.15.021083}
}

@article{han2024electrical,
  title={Electrical 180 switching of {N}{\'e}el vector in spin-splitting antiferromagnet},
  author={Han, Lei and Fu, Xizhi and Peng, Rui and Cheng, Xingkai and Dai, Jiankun and Liu, Liangyang and Li, Yidian and Zhang, Yichi and Zhu, Wenxuan and Bai, Hua and others},
  journal={Sci. Adv.},
  volume={10},
  number={4},
  pages={eadn0479},
  year={2024},
  publisher={American Association for the Advancement of Science},
  doi={10.1126/sciadv.adn0479},
  url={https://doi.org/10.1126/sciadv.adn0479}
}

@article{zhou2025manipulation,
  title={Manipulation of the altermagnetic order in {C}r{S}b via crystal symmetry},
  author={Zhou, Zhiyuan and Cheng, Xingkai and Hu, Mengli and Chu, Ruiyue and Bai, Hua and Han, Lei and Liu, Junwei and Pan, Feng and Song, Cheng},
  journal={Nature},
  volume={638},
  number={8051},
  pages={645--650},
  year={2025},
  publisher={Nature Publishing Group UK London},
  doi={10.1038/s41586-024-08436-3},
  url={https://doi.org/10.1038/s41586-024-08436-3}
}

@article{song2025altermagnets,
  title={Altermagnets as a new class of functional materials},
  author={Song, Cheng and Bai, Hua and Zhou, Zhiyuan and Han, Lei and Reichlova, Helena and Dil, J Hugo and Liu, Junwei and Chen, Xianzhe and Pan, Feng},
  journal={Nat. Rev. Mater.},
  volume={10},
  number={6},
  pages={473--485},
  year={2025},
  publisher={Nature Publishing Group UK London},
  doi={10.1038/s41578-025-00779-1},
  url={https://doi.org/10.1038/s41578-025-00779-1}
}

@article{bai2024altermagnetism,
  title={Altermagnetism: {E}xploring new frontiers in magnetism and spintronics},
  author={Bai, Ling and Feng, Wanxiang and Liu, Siyuan and {\v{S}}mejkal, Libor and Mokrousov, Yuriy and Yao, Yugui},
  journal={Adv. Funct. Mater.},
  volume={34},
  number={49},
  pages={2409327},
  year={2024},
  publisher={Wiley Online Library},
  doi={10.1002/adfm.202409327},
  url={https://doi.org/10.1002/adfm.202409327}
}

@article{jungwirth2026altermagnetic,
  title={Altermagnetic spintronics},
  author={Jungwirth, Tomas and Sinova, J and Wadley, P and Kriegner, D and Reichlova, H and Krizek, F and Ohno, H and {\v{S}}mejkal, L},
  journal={Nature Physics},
  pages={1--10},
  year={2026},
  publisher={Nature Publishing Group UK London},
  doi={10.1038/s41567-026-03337-w},
  url={https://doi.org/10.1038/s41567-026-03337-w}
}

@article{liu2025physical,
  title={Physical properties and first-principles calculations of an altermagnet candidate {C}s\(_{1-\delta}\){V}\(_{2}\){T}e\(_{2}\){O}},
  author={Liu, Chang-Chao and Li, Jing and Liu, Ji-Yong and Lu, Jia-Yi and Li, Hua-Xun and Liu, Yi and Cao, Guang-Han},
  journal={Phys. Rev. B},
  volume={112},
  number={22},
  pages={224439},
  year={2025},
  publisher={APS},
  doi={10.1103/vch1-4khc},
  url={https://link.aps.org/doi/10.1103/vch1-4khc}
}

@article{jiang2025metallic,
  title={A metallic room-temperature \(d\)-wave altermagnet},
  author={Jiang, Bei and Hu, Mingzhe and Bai, Jianli and Song, Ziyin and Mu, Chao and Qu, Gexing and Li, Wan and Zhu, Wenliang and Pi, Hanqi and Wei, Zhongxu and others},
  journal={Nat. Phys.},
  volume={21},
  number={5},
  pages={754--759},
  year={2025},
  publisher={Nature Publishing Group UK London},
  doi={10.1038/s41567-025-02822-y},
  url={https://doi.org/10.1038/s41567-025-02822-y}
}

@article{zhang2025crystal,
  title={Crystal-symmetry-paired spin--valley locking in a layered room-temperature metallic altermagnet candidate},
  author={Zhang, Fayuan and Cheng, Xingkai and Yin, Zhouyi and Liu, Changchao and Deng, Liwei and Qiao, Yuxi and Shi, Zheng and Zhang, Shuxuan and Lin, Junhao and Liu, Zhengtai and others},
  journal={Nat. Phys.},
  volume={21},
  number={5},
  pages={760--767},
  year={2025},
  publisher={Nature Publishing Group UK London},
  doi={10.1038/s41567-025-02823-x},
  url={https://doi.org/10.1038/s41567-025-02823-x}
}

@article{PhysRevLett.63.2144,
  title = {{\ensuremath{\eta} pairing and off-diagonal long-range order in a Hubbard model}},
  author = {Yang, Chen Ning},
  journal = {Phys. Rev. Lett.},
  volume = {63},
  issue = {19},
  pages = {2144--2147},
  numpages = {0},
  year = {1989},
  month = {Nov},
  publisher = {American Physical Society},
  doi = {10.1103/PhysRevLett.63.2144},
  url = {https://link.aps.org/doi/10.1103/PhysRevLett.63.2144}
}

@article{sun2026emergent,
  title={Emergent superconductivity at 16.3 {K} in an altermagnetic candidate {N}a\(_{2-x}\){V}\(_{2}\){S}e\(_{2}\){O} with broken inversion symmetry},
  author={Sun, Yingrui and Yin, Zhaocheng and Zhang, Tengfeng and Wang, Lihan and Ruan, Binbin and Huang, Yu and He, Junbao and Zhu, Wenliang and Ma, Maoyun and Bai, Jianli and others},
  journal={arXiv:2604.00838},
  year={2026},
  url={https://arxiv.org/abs/2604.00838}
}

@article{li2026pressure,
  title={Pressure-induced superconducting-like transition in the altermagnet candidate {C}s{V}\(_{2}\){S}e\(_{2}\){O}},
  author={Li, Yuanzhe and Han, Yilin and Yang, Liu and He, Wanli and Ye, Pengda and Huang, Wencheng and Qiao, Jiabin and Li, Yuemei and Sun, Xiaodong and He, Tingli and others},
  journal={arXiv:2604.09457},
  year={2026},
  url={https://arxiv.org/abs/2604.09457}
}

@article{Jiang2025,
  author = {Jiang, B. and Hu, M. and Bai, J. and Song, Z. and Mu, C. and Qu, G. and Li, W. and Zhu, W. and Pi, H. and Wei, Z. and Sun, Y.-J. and Huang, Y. and Zheng, X. and Peng, Y. and He, L. and Li, S. and Luo, J. and Li, Z. and Chen, G. and Li, H. and Weng, H. and Qian, T.},
  title = {A metallic room-temperature $d$-wave altermagnet},
  journal = {Nat. Phys.},
  volume = {21},
  number = {5},
  pages = {754--759},
  year = {2025},
  doi = {10.1038/s41567-025-02822-y}
}

@article{bf1n-sxdl,
  title = {$d$-Wave Flat Fermi Surface in Altermagnets Enables Maximum Charge-to-Spin Conversion},
  author = {Lai, Junwen and Yu, Tianye and Liu, Peitao and Liu, Long and Xing, Guozhong and Chen, Xing-Qiu and Sun, Yan},
  journal = {Phys. Rev. Lett.},
  volume = {135},
  issue = {25},
  pages = {256702},
  numpages = {7},
  year = {2025},
  month = {Dec},
  publisher = {American Physical Society},
  doi = {10.1103/bf1n-sxdl},
  url = {https://link.aps.org/doi/10.1103/bf1n-sxdl}
}

@article{r8nc-dpt8,
  title = {Electronic structure, magnetic transition, and Fermi surface instability of the room-temperature altermagnet {${\mathrm{KV}}_{2}{\mathrm{Se}}_{2}\mathrm{O}$}},
  author = {Xu, Yuanji and Zhang, Huiyuan and Feng, Maoyuan and Tian, Fuyang},
  journal = {Phys. Rev. B},
  volume = {112},
  issue = {12},
  pages = {125141},
  numpages = {7},
  year = {2025},
  month = {Sep},
  publisher = {American Physical Society},
  doi = {10.1103/r8nc-dpt8},
  url = {https://link.aps.org/doi/10.1103/r8nc-dpt8}
}

@article{yang2025altermagnetic,
author = {X. Yang and S. Fang and Z. Yang and P. Ho and J. Lu and Y. S. Ang},
title = {Altermagnetic Flatband-Driven Fermi Surface Geometry for Giant Tunneling Magnetoresistance},
journal = {Adv. Funct. Mater.},
year = {2026},
volume = {36},
number = {35},
pages = {e31921},
doi = {10.1002/adfm.202531921}
}

@article{yan2025sdw,
  author  = {Yan, Xu and Song, Ziyin and Song, Juntao and Fang, Zhong and Weng, Hongming and Wu, Quansheng},
  title   = {{Magnetic symmetry breaking driven ``inverse magnetic breakdown'' in a d-wave altermagnet KV$_2$Se$_2$O}},
  journal = {Sci. China Phys. Mech. Astron.},
  volume  = {69},
  number  = {5},
  pages   = {257011},
  year    = {2026},
  doi     = {10.1007/s11433-025-2881-5}
}

@article{yin2026discovery,
  title={Discovery of a symmetry-driven electronic cascade in a $d$-wave altermagnet},
  author={Yin, Zhouyi and Shi, Zheng and Zhang, Shuxuan and Liu, Changchao and Cheng, Xingkai and Zhang, Fayuan and Deng, Han-Bin and Yin, Jia-Xin and Chen, Chaoyu and Cao, Guang-Han and others},
  journal={arXiv:2607.11567},
  year={2026},
  url={https://doi.org/10.48550/arXiv.2607.11567}
}

@article{reid2012d,
  title={From $d$-wave to $s$-wave pairing in the iron-pnictide superconductor {(Ba, K)} {F}e$_2${A}s$_2$},
  author={Reid, J-Ph and Juneau-Fecteau, A and Gordon, RT and Ren{\'e} de Cotret, S and Doiron-Leyraud, N and Luo, XG and Shakeripour, H and Chang, J and Tanatar, Makariy A and Kim, Hyunsoo and others},
  journal={Supercond. Sci. and Technol.},
  volume={25},
  number={8},
  pages={084013},
  year={2012},
  publisher={IOP Publishing},
  doi={10.1088/0953-2048/25/8/084013},
  url={https://iopscience.iop.org/article/10.1088/0953-2048/25/8/084013}
}

@article{yang2025observation,
  title={Observation of hidden altermagnetism in {C}s$_{1-\delta}${V}$_2${T}e$_{2}${O}},
  author={Yang, Guowei and Chen, Ruihan and Liu, Changchao and Li, Jing and Pan, Ze and Deng, Liwei and Zheng, Naifu and Tang, Yu and Zheng, Hao and Zhu, Weifan and others},
  journal={arXiv preprint arXiv:2512.00972},
  year={2025},
  url={https://doi.org/10.48550/arXiv.2512.00972}
}

@article{mukasa2025finite,
  title={Finite-momentum superconductivity in two-dimensional altermagnets with a {R}ashba-type spin--orbit coupling},
  author={Mukasa, Kohei and Masaki, Yusuke},
  journal={Journal of the Physical Society of Japan},
  volume={94},
  number={6},
  pages={064705},
  year={2025},
  publisher={The Physical Society of Japan},
  doi={10.7566/JPSJ.94.064705},
  url={https://journals.jps.jp/doi/10.7566/JPSJ.94.064705}
}

@article{3k12-2467,
  title = {Phase-modulated superconductivity via altermagnetism},
  author = {Sumita, Shuntaro and Naka, Makoto and Seo, Hitoshi},
  journal = {Phys. Rev. B},
  volume = {112},
  issue = {14},
  pages = {144510},
  numpages = {18},
  year = {2025},
  month = {Oct},
  publisher = {American Physical Society},
  doi = {10.1103/3k12-2467},
  url = {https://link.aps.org/doi/10.1103/3k12-2467}
}

@article{cvq6-b1nf,
  title = {Fulde-Ferrell-Larkin-Ovchinnikov states and topological Bogoliubov Fermi surfaces in altermagnets: An analytical study},
  author = {Liu, Zhao and Hu, Hui and Liu, Xia-Ji},
  journal = {Phys. Rev. B},
  volume = {113},
  issue = {2},
  pages = {024518},
  numpages = {15},
  year = {2026},
  month = {Jan},
  publisher = {American Physical Society},
  doi = {10.1103/cvq6-b1nf},
  url = {https://link.aps.org/doi/10.1103/cvq6-b1nf}
}

@article{yang2026emergent,
  title={Emergent spin-resolved electronic charge density waves and pseudogap phenomena from strong $d$-wave altermagnetism},
  author={Yang, Fei and Zhao, Guo-Dong and Yan, Binghai and Chen, Long-Qing},
  journal={arXiv:2602.11694},
  year={2026}
}

@article{gu2026real,
  title={Real-space identification of distinct magnetic configurations in a candidate d-wave altermagnet},
  author={Gu, Jin-Cheng and Hu, Mingzhe and Song, Ziyin and Wang, Lihan and Wang, Lihong and Zhang, Junming and Zhao, Jiali and Li, Hang and Jin, Shifeng and Zhang, Xin-Ding and others},
  journal={arXiv preprint arXiv:2606.29140},
  year={2026},
  url={https://doi.org/10.48550/arXiv.2606.29140}
}

@article{sun2025antiferromagnetic,
  title = {Antiferromagnetic structure of ${\mathrm{KV}}_{2}{\mathrm{Se}}_{2}\mathrm{O}$: A neutron diffraction study},
  author = {Sun, Yili and Huang, Yu and Cheng, Jingwen and Zhang, Shanshan and Li, Zezhong and Luo, Huiqian and Ma, Xiaobai and Yang, Wenyun and Yang, Jinbo and Chen, Dongfeng and Sun, Kai and Gutmann, Matthias and Capelli, Silvia C. and Shen, Feiran and Hao, Jiazheng and He, Lunhua and Chen, Genfu and Li, Shiliang},
  journal = {Phys. Rev. B},
  volume = {112},
  issue = {18},
  pages = {184416},
  numpages = {6},
  year = {2025},
  month = {Nov},
  publisher = {American Physical Society},
  doi = {10.1103/27n8-5q4l},
  url = {https://link.aps.org/doi/10.1103/27n8-5q4l}
}

\end{document}